\documentclass[12pt]{article} 
\usepackage{amsmath,amssymb,subfigure,amsthm}
\usepackage[dvips]{epsfig}
\usepackage{amsmath,amssymb,subfigure}
\usepackage[dvips]{epsfig}
\makeatletter \@addtoreset{equation}{section} \makeatother
\renewcommand{\theequation}{\arabic{section}.\arabic{equation}}
\begin{document}
\pagestyle{empty}
\newtheorem*{pb}{Property B.1}
\newtheorem{theo}{Theorem}[section] 
\newtheorem{lem}[theo]{Lemma}
\newtheorem{cor}[theo]{Corollary}
\newtheorem{pro}[theo]{Proposition}
\newtheorem{prop}{Property}
\newcommand{\no}{\nonumber}
\newcommand{\p}{\partial}
\newcommand{\lb}{\lambda}
\newcommand{\vu}{\vec{u}}
\newcommand{\tvu}{\vec{\tilde{u}}}
\newcommand{\e}{\epsilon}
\newcommand{\ra}{\rightarrow}

\begin{center}
{\large\bf Large Parameter Behavior of Equilibrium Measures}
\end{center}

\vspace*{.1in}

\begin{center}
Tamara Grava
\vspace{.1in}

{\em Scuola Internazionale Superiore di Studi Avanzati, via Beirut 4, 34014 Trieste, Italy\\
grava@sissa.it }

and

Fei-Ran Tian
\vspace{.1in}

{\em Department of Mathematics, Ohio State University, Columbus, OH 43210 \\
tian@math.ohio-state.edu}

\end{center}

\vspace*{.5in}

\noindent
{\bf Abstract:} \quad
We study the equilibrium measure for a logarithmic potential in the presence of an 
external field  $V_*(\xi) + t \ p(\xi)$, where $t$ is a parameter, $V_*(\xi)$ is a smooth function and 
$p(\xi)$ a monic polynomial. When $p(\xi)$ is of an odd degree, the equilibrium measure is shown to
be supported on 
a single interval as $|t|$ is sufficiently large.
When $p(\xi)$ is of an
even degree, the equilibrium measure is supported on two disjoint intervals as $t$ is negatively
large; it is supported on a single interval for convex $p(x)$ as $t$ is positively large and
is likely to be supported on multiple disjoint intervals for non-convex $p(x)$.

The support of the equilibrium measure shrinks to isolated points as $|t| \rightarrow +\infty$
in all the cases that we consider. For sufficiently large $|t|$, each topological component of 
the support contains
a local minimizing point of the external field $V_*(\xi) + t \ p(\xi)$; a ``potential well''
phenomenon.

\thispagestyle{empty}

\newpage

\setcounter{page}{1}
\pagestyle{plain}
\refstepcounter{section}
\begin{center}
{\bf $\S$ 1 \quad Introduction}
\end{center}

In this paper, we study the following minimization problem with constraints
\begin{equation}
\label{mini}
\underset{\{\psi \geq 0,\,\smallint \psi
d\xi=1\}}{\rm
Minimize}\left[
-\frac{1}{2\pi}\int_{- \infty}^{+\infty}\int_{- \infty}^{+\infty} \log|\xi-\eta|\psi(\xi)\psi(\eta)d\xi
d\eta + 
\int_{- \infty}^{+\infty}
V(\xi)\psi(\xi)d\xi\right]
\ .
\end{equation}
The external field $V(\xi)$ is a $C^{\infty}$ function that 
satisfies
\begin{equation}
\label{infty}
\lim_{\xi \rightarrow \pm \infty} \dfrac{V(\xi)}{log(1 + \xi^2)} = + \infty \ .
\end{equation}
Under this condition, the 
existence and uniqueness of the minimizer for (\ref{mini})
has been established \cite{saf}. The measure $\psi(\xi) d \xi$, where 
$\psi(\xi)$ is the minimizer of (\ref{mini}), is called the equilibrium measure
under the external field $V(\xi)$. 

Equilibrium measures find applications in many 
branches of mathematical sciences.
It is used to describe the partition function of the Hermitian 
one-matrix model in  random matrix theory of statistical 
physics \cite{bre,meh}.
It is also intrinsically connected
to the free energy of the Yang-Mills theory \cite{dub}. Finally, 
it plays an important role in orthogonal polynomials 
and approximation theory \cite{saf}.

Although its
importance to physics and
approximation theory
(\cite{dij, saf}) is well known, the minimization
problem (\ref{mini}) is not well
understood. The minimizer is
explicitly known only for a few cases where the external fields
$V(\xi)$ are the simplest
polynomials \cite{Zu, dei1, saf}. 
It is therefore desirable to
study the minimization
problem for much more general  $C^{\infty}$ external fields.

Our method to solve the minimization problem (\ref{mini}) is the same as
one \cite{GT} we used to solve a similar minimization problem 
for the zero dispersion limit of the KdV equation
$$u_t + 6 u u_x + \epsilon^2 u_{xxx} = 0  \quad 
\mbox{with $u(x, 0; \epsilon) = u_0(x)$} \ ,$$ 
where the initial data $u_0(x)$ is a bounded decreasing function.
The weak limit of the KdV solution $u(x,t;\epsilon)$ as $\epsilon \rightarrow 0$ is 
determined by a minimization problem \cite{lax, lax2, ven}
\begin{equation}
\label{min}
\underset{\{\psi \geq 0, \  \psi \in L^1 \}}
{\rm
Minimize} \left[- \frac{1}{2 \pi} \iint \log \Big|\frac{\xi - \eta}
{\xi + \eta
}\Big|
\psi(\xi) \psi(\eta) d \xi d \eta + \int V(\xi, x, t)
\psi(\xi) d
\xi
\right] \ ,
\end{equation}
where the space-time dependence is through the
function $\tilde{V}$
\begin{equation}
\label{V'}
\tilde{V}(\xi, x, t) = x \xi - 4 \xi^3 t
- \theta(\xi) 
\end{equation}
and
$\theta(\xi)$ is encoded with the initial
information $u_0(x)$. We found that the problem (\ref{min}) is intrinsically 
connected to the Euler-Poisson-Darboux equations \cite{GT}. We utilized the solution of 
the equations to study the minimizer and hence the weak limit of 
the KdV solution.

The minimizer of (\ref{mini}) has been known to have a
compact support
if the external field $V(\xi)$ satisfies condition (\ref{infty}). 
The support
is usually a union of a finite or infinite number of disjoint closed intervals.
In this paper, we will study how the number of gaps in the support
of the minimizer varies with respect to the external field.
We would like to know whether there is
any universality on the number of gaps. 

We shall consider parameter-dependent
external field $V(\xi) = V_*(\xi) + t \ p(\xi)$, where $t$ is a parameter,
$V_*(\xi)$
is a smooth function of $\xi$ and $p(\xi)$ a monic polynomial. This is motivated by 
the space-time dependence of $\tilde{V}$ of (\ref{V'}) in the KdV minimization
problem (\ref{min}).
We shall show that when $p(\xi)$ is of an odd degree, the minimizer of (\ref{mini}) has
no gap in its compact support as $|t|$ is sufficiently large.
When $p(\xi)$ is of an even degree, the minimizer has a single gap in its
support as $t$ is negatively
large; it has no gap for convex $p(\xi)$ as $t$ is positively large and is likely to have
multi-gaps for non-convex $p(\xi)$.
We note that the same minimizer
may have an arbitrary number of gaps when $|t|$ is not large.
A similar result has also been discovered in the case of the KdV weak limit \cite{Tian2}. 
Namely, the KdV weak
limit is of zero or single phase for all $x$ when $t$ is sufficiently large, i.e.,
the minimizer 
of (\ref{mini}) has either zero or one gap in its support.

The support of the minimizer shrinks to isolated points as $|t| \rightarrow +\infty$
in all the cases that we consider. Each topological component of the support contains
a local minimizing point of the external field $V(\xi)$; a ``potential well''
phenomenon.

The organization of the paper is as follows.

In Section 2, we will use function theoretical methods to solve the 
minimization  problem. We will formulate the minimizer in terms of solutions 
of the  Euler-Poisson-Darboux equations.

In Section 3, we will use the Euler-Poisson-Darboux solutions to study the behavior of 
the minimizer when the parameter
in the external field is sufficiently large.

\refstepcounter{section}
\begin{center}
{\bf $\S$ 2 \quad Solution of the Minimization Problem}
\end{center}

In this section, we will use the method that we have developed for the KdV
zero dispersion limit \cite{GT} to solve the minimization problem (\ref{mini}).

Introducing a linear operator
\begin{equation}
L \psi (\xi) = \dfrac{1}{\pi} \int_{-\infty}^{+\infty} \log |\xi - \mu|
\psi(\mu) d \mu \ , \label{L}
\end{equation}
we rewrite the quadratic functional of (\ref{mini})
as  $ - \dfrac{1}{2} <L \psi, \psi> + <V, \psi> $.
Here $<~>$ is the standard $L^2$ inner product.
The  Euler-Lagrange equations take the form 
\begin{eqnarray}
L \psi (\xi) - V(\xi) &=& l \quad \mbox{where
$\psi > 0$}  \ , \label{L1} \\
L \psi (\xi) - V(\xi) &\leq& l \quad \mbox{where
$\psi =0$} \ , \label{L2}
\end{eqnarray}
where $l$ is the Lagrange multiplier. 
It can be shown \cite{dei0} that $\psi$ is the minimizer 
iff $\psi$ is
a nonnegative function that satisfies variational
conditions (\ref{L1}-\ref{L2}) and the constraint
\begin{equation}
\int_{- \infty}^{+\infty} \psi(\xi) d \xi = 1 \ . \label{con}
\end{equation}

We make the ansatz that the support of $\psi$ consists
of a finite union of disjoint intervals. One denotes 
$I = \{ \xi; ~ \psi > 0 \}$ and writes
\begin{equation}
\label{support}
I = \cup_{k=1}^{g+1} (u_{2k}, u_{2k-1}) \ ,
\end{equation}
where $u_{2g+2} < \cdots < u_2 < u_1$. Hence, the support is
the closure of $I$. 

We now consider a slightly stronger version of (\ref{L1}) and (\ref{L2}),
\begin{eqnarray}
L \psi (\xi) - V(\xi) &=& l \quad \mbox{on $I$}
 \ , \label{L3} \\
L \psi (\xi) - V(\xi) &<& l \quad \mbox{on $\mathbb{R} \backslash \bar{I}$}
\ , \label{L4}
\end{eqnarray}
where $\bar{I}$ denotes the closure of $I$. Since $\psi \geq 0$,
we must also have
\begin{eqnarray}
\psi &>& 0  \quad \mbox{on $I$} \ , \label{L5} \\
\psi &=& 0  \quad \mbox{off $I$} \ . \label{L6}
\end{eqnarray}

Our strategy to construct the minimizer is to first find the 
solution $\psi$ of equations (\ref{con}),
(\ref{L3}) and (\ref{L6}) and then impose inequalities (\ref{L4}) and (\ref{L5}) on $\psi$.
Since it is a non-negative function and since it satisfies (\ref{L1}), 
(\ref{L2}) and (\ref{con}), $\psi$ will then be the minimizer.

To solve (\ref{con}), (\ref{L3}) and (\ref{L6}), we use the fact  that the operator $L$ of 
(\ref{L}) is connected to the Hilbert transform $H$ on the real line.
\begin{equation}
\label{lh}
L \psi (\xi) = \int_0^{\xi} H \psi (\tau) d \tau \ ,
\end{equation}
where
\begin{displaymath}
H \psi (\xi) = {1 \over \pi} P.V. \int_{- \infty}^{\infty}
{\psi(\mu)  d \mu \over \xi - \mu}  \ .
\end{displaymath}
This makes equations (\ref{L3}) and (\ref{L6}) amenable to the 
Riemann-Hilbert technique in function theory.

Differentiating (\ref{L3}) with respect to $\xi$ and using (\ref{lh}),
one obtains
\begin{equation}
H \psi (\xi) = V'(\xi)  ~~~~~ \mbox{on I} \ .
\label{H1}
\end{equation}
To recover equation (\ref{L3}) from
(\ref{H1}) by integration, one must have
for $k$ $=$ $1$, $2$, $\cdots$, $g$,
\begin{equation}
\label{loop'}
\int_{u_{2k+1}}^{u_{2k}} [H \psi(\xi) - V'(\xi)] d \xi = 0 \ .
\end{equation}

Recalling the relation of the Hilbert transform to analytic function, one
can write for real $\xi$
\begin{displaymath}
{\cal G}^+(\xi) = \psi(\xi) + \sqrt{-1} H \psi (\xi) \ ,
\end{displaymath}
where ${\cal G}^+(\xi)$ is the boundary value on the real axis of a
function
$${\cal G}(z) = {1 \over \pi \sqrt{-1}} \int_{- \infty}^{+ \infty} {\psi(\mu) d \mu 
\over \mu - z} \ ,$$ 
which is analytic in the upper half complex plane.
In view of the constraint (\ref{con}), we expand the Cauchy integral to obtain
${\cal G}(z) = - {1 \over \sqrt{-1} \pi z} + O(1/z^2)$ for large $|z|$.

Conditions (\ref{loop'}) now take a new form
\begin{equation}
\label{loop}
\int_{u_{2k+1}}^{u_{2k}} [Im {\cal G}^+(\xi) - V'(\xi)] d \xi = 0 
\end{equation}
for $k=1, 2, \cdots, g$.

Equations (\ref{L6}) and (\ref{H1}) then
become a Riemann-Hilbert
problem in function theory
\begin{eqnarray}
Im {\cal G}^+(\xi) &=& V'(\xi) ~~~~~ \mbox{on $I$} \ , \label{rh1} \\
Re {\cal G}^+(\xi) &=& 0 ~~~~~~~~~~\mbox{off $I$} \ , \label{rh2}
\end{eqnarray}
where ${\cal G}(z)$ is analytic in the upper
half complex plane.
It follows from the Plemelj formula that
\begin{displaymath}
{\cal \tilde{G}}(z) = {R(z,\vec{u}) \over \pi} \int_{I} {V'(\mu) \over
R(\mu, \vec{u}) (\mu - z)} d \mu \ ,
\end{displaymath}
where $\vec{u}$ denotes $(u_1, u_2, \cdots, u_{2g+2})$
and $R(\xi, \vec{u}) = \sqrt{(\xi - u_1)(\xi - u_2) \cdots (\xi - u_{2g+2})}$,
is a solution to this Riemann-Hilbert problem.
Here $R(\xi, \vec{u})$ is set to be positive 
for $\xi> u_1$. It defines a Riemann surface with branch cuts along the set 
$I$ of (\ref{support}).

To derive the equations governing the 
endpoints $u_1$, $u_2$, $\cdots$, $u_{2g+2}$,
one usually imposes 
condition (\ref{loop}) and the asymptotics ${\cal G}(z) = - {1 \over \sqrt{-1} \pi z} + O(1/z^2)$ for 
large $|z|$ \cite{dei1, jur}.
The former gives $g$ equations and the latter results in $g+2$ moment conditions.
Consequently, one obtains
exactly $2g+2$ algebraic equations on $u_1$, $u_2$,
$\cdots$, $u_{2g+2}$. 

In this paper, we take a slightly different approach.
We observe
that the Riemann-Hilbert problem (\ref{rh1}-\ref{rh2}) 
has many other solutions. Indeed,
it is obvious that
\begin{equation}
\label{g}
{\cal G}(z) =  {  {R^2(z, \vec{u}) \over \pi}
\int_{I} {V'(\mu) \over
R(\mu, \vec{u}) (\mu - z)} d \mu + \sqrt{-1}
Q(z) \over
R(z, \vec{u}) } \ ,
\end{equation}
where $Q(z)$ is an arbitrary polynomial
with real coefficients, is also a solution.
We choose polynomial $Q(z)$ such that 

\begin{enumerate}
\item it is a polynomial of degree $2g+1$.

\item ${\cal G}(z) = - {1 \over \sqrt{-1} \pi z} + O(1/z^2)$ for 
large $|z|$.

\item conditions (\ref{loop}) are satisfied.

\end{enumerate}

Hence, constraint (\ref{con}) and conditions 
(\ref{L3}) and (\ref{L6}) are built in the construction of ${\cal G}$.

It is quite obvious that such a polynomial $Q$ is unique.

We first analyze the boundary value of
${\cal G}(z)$ on the real axis. Necessarily, the Cauchy integral
in the numerator of (\ref{g}) will become a singular integral.
Our key observation is that the latter is, more or less, the solution of a
boundary value problem for
the Euler-Poisson-Darboux equations.

\begin{pro}
\label{prop0}
The boundary value of ${\cal G}(z)$ at $\xi \neq u_i$, $i=1, 2, \cdots,
2g+2$
\begin{equation}
\label{g2}
{\cal G}^+(\xi)={- 2 \sqrt{-1} R^2(\xi, \vec{u})
\Phi_g(\xi, u_1, u_2, \cdots, u_{2g+1}) + \sqrt{-1} Q(\xi) \over
R(\xi, \vec{u})} + \sqrt{-1} V'(\xi) \ ,
\end{equation}
where $\Phi_g(\xi, u_1, \cdots, u_{2g+2})$ satisfies the 
Euler-Poisson-Darboux equations
\begin{eqnarray}
2(u_i - u_j) {\p^2 \Phi_g \over \p u_i \p u_j} &=& 
{\p \Phi_g \over \p u_i}
- {\p \Phi_g \over \p u_j} \ , \label{ph1} \\
2(\xi - u_i) {\p^2 \Phi_g \over \p \xi \p u_i} &=& 
{\p \Phi_g \over \p \xi}
- 2 {\p \Phi_g \over \p u_i} \ , \label{ph2} \\
\Phi_g(u, u, \cdots, u) &=&
{1 \over 2 (g+1)!} {d^{g+2} V(u) \over d u^{g+2}} \ . \label{BP}
\end{eqnarray}
\end{pro}

We will omit the proof here, since it is similar to the proof of an
analogous result on the KdV zero dispersion limit \cite{GT}.

A formula for $Q$ can be derived from the loop conditions (\ref{loop})
and asymptotics ${\cal G}(z) = - {1 \over \sqrt{-1} \pi z} + O(1/z^2)$ for 
large $|z|$. To achieve this, we introduce a sequence of polynomials
\begin{equation}
\label{pgn}
P_{g,n}(\xi,\vec{u}) = \xi^{g+n} + a_{g,1} \xi^{g+n-1} + \cdots + 
a_{g, g+n}
\end{equation}
whose coefficients are uniquely determined by
\begin{equation}
\label{Pgn}
\frac{P_{g,n}(\xi,\vec{u})}{R(\xi, \vu)} =
\xi^{n-1} +
O({1 \over \xi^2})  ~~~~~~~
\mbox{for large $|\xi|$} \ ,
\end{equation}
and
\begin{equation}
\label{lc}
\int_{u_{2k+1}}^{u_{2k}} \frac{P_{g,n}(\xi,\vec{u})}{R(\xi, \vu)} d \xi
= 0  ~~~~~~~~
k = 1, 2, \cdots, g \ .
\end{equation}

\begin{pro}
\label{prop1}
\begin{eqnarray*}
Q(\xi, \vu) &=&  -\sum_{i = 1}^{2g+2}
[\prod_{l=1, l \neq i}^{2g+2} (\xi-u_l)] \Psi_g(u_i, \vu)
+ {1 \over \pi} P_{g, 0}(\xi, \vu) + 
c_1 P_{g, 1}(\xi, \vu) \\
&& + \cdots + c_{g} P_{g, g}(\xi, \vu) \ . \label{q}
\end{eqnarray*}
Here $\Psi_g(\xi, u_1, u_2, \cdots, u_{2g+2})$ satisfies the 
Euler-Poisson-Darboux equations 
(\ref{ph1}) and 
(\ref{ph2}) with
the diagonal boundary value
\begin{equation}
\Psi_g(u, u, \cdots, u) = {1 \over 2 (g+1)!} {d^{g+1} V(u) 
\over d u^{g+1}} \ .
\label{BP2}
\end{equation}
The coefficients
$$c_k = 2k \sum_{l=0}^{g-k} \Gamma_l(\vu) q_{g, k+l}(\vu) ~~~~~~~~ k=1,2, \cdots, g, $$
where $\Gamma_l(\vu)$'s come from the expansion
\begin{equation}
\label{gamma}
R(\mu, \vu) = \mu^{g+1} [\Gamma_0(\vu) + {\Gamma_1(\vu) \over \mu} + 
{\Gamma_2(\vu) \over \mu^2} + \cdots ] \ .
\end{equation}
The function $q_{g, k}$ is 
\begin{equation}
\label{qk}
q_{g, k}(\vu) = {1 \over 2 \pi \sqrt{-1}} \int_{I} {V(\mu) \mu^{g - k} \over R(\mu, \vu)} \ d \mu \ .    
\end{equation}
\end{pro}
The proof is also similar to that of the KdV case \cite{GT}.

The function $q_{g,k}(\vu)$ also satisfies equations of Euler-Poisson-Darboux type \cite{GT}.

Finally, we postulate the boundedness of the minimizer, which is equal 
to the real part of $\cal G^+(\xi)$.
The numerator in (\ref{g2}) must then vanish at the end points of set $I$
\begin{equation}
\label{Hodo2}
P(u_i, \vu) = 0  ~~~~~ \quad i=1, 2, \cdots, 2g+2 \ ,
\end{equation}
where
\begin{equation}
\label{P}
P(\xi, \vu) = 2 R^2(\xi, \vu) \Phi_g(\xi, \vu) - Q(\xi, \vu) \ .
\end{equation}
Since it is of degree $2g+1$ in $\xi$ and has $2g+2$ zeros because
of (\ref{Hodo2}), the  polynomial $Q(\xi, \vu)$ must be identically
zero. The real part of ${\cal G}^+$ of (\ref{g2}), on the solution 
$\vec{u}$
of equations (\ref{Hodo2}), becomes
\begin{equation}
\label{minimizer}
\psi(\xi) = - 2  Re\{\sqrt{-1}R(\xi, \vu)\}
\Phi_g(\xi, \vu) \ .
\end{equation}
Hence, $\psi(\xi)$ of (\ref{minimizer}) satisfies
conditions (\ref{con}), (\ref{L3}) and (\ref{L6}). 

Inequality (\ref{L5}) is equivalent to 
\begin{equation}
Re\{\sqrt{-1}R(\xi,\vu)\}\Phi_g(\xi,
\vu) < 0  \quad
\mbox{for $\xi$ on I} \ , \label{constraint} 
\end{equation}
and (\ref{L4}) is equivalent to
\begin{eqnarray}
\int_{u_{2k+1}}^{\xi} R(\mu, \vu) \Phi_g(\mu,\vu)
d \mu & > & 0  \quad \mbox{for
$u_{2k+1}<\xi<u_{2k}$ and $1 \leq k \leq 2g+2$} \ , \no \\
\int_{u_1}^{\xi} R(\mu, \vu) \Phi_g(\mu,\vu)
d \mu & > & 0  \quad \mbox{for $\xi > u_1$} \ , \label{constraint2} \\
\int_{\xi}^{u_{2g+2}} R(\mu, \vu) \Phi_g(\mu,\vu)
d \mu & < & 0  \quad \mbox{for $\xi < u_{2g+2}$} \ . \no
\end{eqnarray}

We summarize the above results in the following theorem

\begin{theo}
If $(u_1, u_2, \cdots, u_{2g+2})$ satisfies
equations (\ref{Hodo2})
and if inequalities (\ref{constraint}-\ref{constraint2})
are satisfied, then $\psi(\xi)$
of (\ref{minimizer}) is a non-negative function
that satisfies variational conditions
(\ref{L1}) and (\ref{L2}) and constraint (\ref{con}); 
so $\psi(\xi)$ must be the minimizer of the minimization problem
(\ref{mini}).
\end{theo}

\bigskip

\refstepcounter{section}
\begin{center}
{\bf $\S$ 3 \quad Large Parameter Results}
\end{center}

In this section, we shall consider a one-parameter family of external
field
\begin{equation}
\label{V}
V(\xi) = V_*(\xi) + t \ p(\xi) \ .
\end{equation}
Here, $t$ is a parameter, $p(\xi)$ is a monic polynomial of degree $n$ and
$V_*(\xi)$ is a
$C^{\infty}(-\infty,+\infty)$
function.
We are interested in the behavior of the equilibrium measure when
$|t|$ is large.

Our strategy for studying the large parameter behavior of the minimizer
of (\ref{mini}) is as follows.
For simplicity, we will assume $V_*(\xi)$ to have a power function 
growth at $\xi = - \infty$ or $\xi = + \infty$. This allows us to use the scaling
technique to study the equilibrium measure when the parameter $t$ is
sufficiently large.

\medskip

\noindent
{\bf $\S$ 3.1 \quad The degree of polynomial $p(\xi)$ is odd}

\medskip

We shall show that the equilibrium measure is supported on a single interval
$[u_2, u_1]$ when $|t|$ is large. This corresponds to
(\ref{support}) for $g=0$.  

We will use Theorem 2.3 to construct the minimizer of (\ref{mini}). Hence, we need to
solve equations (\ref{Hodo2}) for $g=0$ and verify inequalities
(\ref{constraint}-\ref{constraint2}). 

We now consider the algebraic equations (\ref{Hodo2}) for $g=0$.
The function $P(\xi, u_1, u_2)$ of (\ref{P}) is
\begin{displaymath}
2(\xi - u_1)(\xi - u_2)\Phi_0(\xi,u_1, u_2) +
(\xi-u_2)\Psi_0(u_1,u_1,u_2)+(\xi-u_1)\Psi_0(u_2,u_1,u_2)-{1 \over \pi} \ .
\end{displaymath}
Equations (\ref{Hodo2}) exactly become
\begin{eqnarray}
(u_1-u_2) \Psi_0(u_1,u_1,u_2) -
{1 \over \pi} &=&0 \ , \label{Ht11} \\
(u_2-u_1) \Psi_0(u_2,u_1, u_2)-
{1 \over \pi} &=&0 \ , \label{Ht12}
\end{eqnarray}
which are equivalent to
\begin{eqnarray}
(u_1 - u_2) \sqrt{{\p \over \p u_2}\Psi_{0}(u_1,u_1,u_2)}
- \sqrt{{1 \over \pi}} 
&=& 0 \ , \label{Ht13} \\
\Psi_{0}(u_1, u_1, u_2) + \Psi_{0}(u_2,u_1,u_2) &=& 0 \ . \label{Ht14} 
\end{eqnarray}
In the derivation of (\ref{Ht13}), we have used an identity
\begin{equation}
\label{identity}
\Psi_0(u_1, u_1, u_2) - \Psi_0(u_2, u_1, u_2) = 2(u_1 - u_2) {\p \over \p u_2} \Psi_0(u_1, u_1, u_2) \ .
\end{equation}
This is easily verified by calculating 
\begin{eqnarray*}
\lefteqn{{\p \over \p \xi} \left[ 2(\xi - u_2) {\p \over u_2} \Psi_0(\xi, u_1, u_2) - \Psi_0(\xi, u_1, u_2) \right]} \\
&=& 2(\xi - u_2) {\p^2 \over \p u_2 \p \xi} \Psi_0 + 2 {\p \over \p u_2} \Psi_0 - {\p \over \p \xi} \Psi_0 \ .
\end{eqnarray*}
The right hand side vanishes because $\Psi_g$ satisfies the Euler-Poisson-Darboux equation (\ref{ph2}).
The function in the parenthesis is then independent of $\xi$; so we obtain
$$2(\xi - u_2) {\p \over \p u_2} \Psi_0(\xi, u_1, u_2) - \Psi_0(\xi, u_1, u_2) = - \Psi_0(u_2, u_1, u_2) \ .$$
Letting $\xi = u_1$ yields the identity (\ref{identity}).

We will rewrite equation (\ref{Ht14}) in another useful form.
Its left hand side is a function of $u_1$ and $u_2$. Denote
this function by $H(u_1, u_2)$. Since $\Psi_0(\xi, u_1, u_2)$ satisfies
equations (\ref{ph1}-\ref{ph2}), we derive an equation for $H(u_1, u_2)$
$$2(u_1 - u_2) {\p^2 H \over \p u_1 \p u_2} = {\p H \over \p u_1}
- {\p H \over \p u_2} \ .$$
The boundary condition (\ref{BP2}) for $\Psi_0(\xi, u_1, u_2)$ implies
$H(u,u) = V'(u)$. We then use formula (\ref{EPQsS}) to obtain
$$H(u_1, u_2) = {1 \over \pi} \int_{u_2}^{u_1} {V'(\mu) d \mu \over
\sqrt{(u_1 - \mu)(\mu - u_2)}} \ .$$
Hence, equation (\ref{Ht14}) is equivalent to
\begin{equation}
\label{Ht14'}
\int_{u_2}^{u_1} {V'(\mu) d \mu \over \sqrt{(u_1 - \mu)(\mu - u_2)}} = 0 \ . 
\end{equation}

We shall study the case when $t$ is negatively large in detail.
The other case when $t$ is positively large can be handled in 
the same way.

We now make assumptions on $V_*(\xi)$. Since $V(\xi) = V_*(\xi) + t \ p(\xi)$
satisfies the growth condition (\ref{infty}), $V_*(\xi)$ must grow faster
than $p(\xi)$ as $\xi \rightarrow + \infty$ because of $t < 0$. In contrast, we will
make a very mild assumption on $V_*(\xi)$ as $\xi \ll -1$  
\begin{eqnarray}
\label{V*}
V_*(\xi) &=& C_+ \xi^{M_+} + h_+(\xi) ~~~~~~~~~~~~ \mbox{ for $\xi \gg 1$} \ , \\
V_*'(\xi) &\leq& 0 ~~~~~~~~~~~~~~~~~~~~~~~~~~~~~~~~ \mbox{for $\xi \ll -1$} \ .
\label{V*-}
\end{eqnarray}
Here $C_+$ is a positive constant and $M_+$ is a positive constant that is bigger than $n$ and $2$. 
The number $n$ is the degree
of the polynomial $p(\xi)$. Function $h_+(\xi)$ has an order less than $n$. More 
generally, we assume
\begin{equation}
\label{V*h}
\lim_{\xi \rightarrow + \infty} {h_+'''(\xi) \over \xi^{M_+ -3}} = 0  \ .
\end{equation}
This immediately implies
\begin{equation}
\label{V*hh}
\lim_{\xi \rightarrow + \infty} {h_+(\xi) \over \xi^{M_+}} = 0  \ , \quad
\lim_{\xi \rightarrow + \infty} {h_+'(\xi) \over \xi^{M_+ -1}} = 0  \ , \quad
\lim_{\xi \rightarrow + \infty} {h_+''(\xi) \over \xi^{M_+ -2}} = 0  \ .
\end{equation}

We will solve equations (\ref{Ht13}) and (\ref{Ht14}) for large $u_1$ and $u_2$.
This is motivated by the fact that $V(\xi)$ has a critical point between $u_2$
and $u_1$ in view of equation (\ref{Ht14'}) and that the minimizing point 
of $V(\xi) = V_*(\xi) + t \ p(\xi)$ 
moves to $+ \infty$ as $t \rightarrow - \infty$.
 
We will first consider the case $p(\xi) = \xi^n$.

We will split $\Psi_0$ of (\ref{Ht13}-\ref{Ht14}) into simpler terms.
In view of its boundary data (\ref{BP2}) for $g=0$, $\Psi_0$ depends 
linearly on $V$. The 
decomposition of $V= t \xi^n + C_+ \xi^{M_+ - 1} + h_+(\xi)$ allows us to write 
\begin{equation}
\Psi_0 = t \Psi_{\xi^n} + C_+ \Psi_{\xi^{M_+}} + \Psi_{h_+(\xi)} \ .
\label{split}
\end{equation}
In view of the integral formula (\ref{H}), $\Psi_{\xi^n}$ and $\Psi_{\xi^{M_+}}$ 
are homogeneous functions of $(\xi, u_1, u_2)$
of orders $n-1$ and $M_+ -1$, respectively.
 
Substituting (\ref{split}) into equation (\ref{Ht13}) and then dividing it by 
$|t|^{M_+ \over 2(M_+ -n)}$, we use the homogeneity of $\Psi_{\xi^n}$ and $\Psi_{\xi^{M_+}}$
to obtain
\begin{eqnarray}
\lefteqn{(U_1 - U_2) \{ - {\p \over \p U_2} 
\Psi_{\xi^n}(U_1, U_1, U_2) + C_+ 
{\p \over \p U_2} \Psi_{\xi^{M_+}}(U_1, U_1, U_2)} \no \\
&& + \ {1 \over |t|^{M_+ -2 \over M_+ -n}}
{\p \over \p u_2} \Psi_{h_+
(\xi)}(u_1, u_1, u_2) \}^{1 \over 2} 
- {1 \over \sqrt{\pi} |t|^{M_+ \over 2(M_+ -n)}} = 0 \ . \label{Ht15}
\end{eqnarray}
Here $U_1$ and $U_2$ are
\begin{equation}
U_1 = {u_1 \over |t|^{1 \over M_+ -n}} \ , ~~~ U_2 = {u_2 \over |t|^{1 \over M_+ -n}} \ .
\label{U12}
\end{equation}
 
Similarly, substituting (\ref{split}) into equation (\ref{Ht14}) and then dividing it by
$|t|^{M_+ -1 \over M_+ -n}$, we obtain
\begin{eqnarray}
\lefteqn{- \Psi_{\xi^n}(U_1, U_1, U_2) + C_+ \Psi_{\xi^{M_+}}(U_1, U_1, U_2) +  
{1 \over |t|^{M_+ -1 \over M_+ -n}}  \Psi_{h_+(\xi)}(u_1, u_1, u_2)} \label{Ht16}  \\
&& - \Psi_{\xi^n}(U_2, U_1, U_2) + C_+ \Psi_{\xi^{M_+}}(U_2, U_1, U_2) +
{1 \over |t|^{M_+ -1 \over M_+ -n}}  \Psi_{h_+(\xi)}(u_2, u_1, u_2) = 0 \ . 
\no
\end{eqnarray}

It follows from the behavior (\ref{V*h}-\ref{V*hh}) of $h_+(\xi)$ that the three terms 
involving $\Psi_{h_+(\xi)}$ in (\ref{Ht15}-\ref{Ht16}), together with their first derivatives
with respect to $U_1$ and $U_2$,
decay to zero as $t \rightarrow - \infty$ if $U_1$ and $U_2$ of (\ref{U12}) are kept bounded.

We now denote $1/t$ by $T$ and solve equations (\ref{Ht15}) and (\ref{Ht16})
for $U_1$ and $U_2$ as functions of $T$ in the neighborhood of $T=0$. 

First, equations (\ref{Ht15}) and (\ref{Ht16}) at $T=0$ become
\begin{eqnarray}
(U_1 - U_2) \sqrt{- {\p \over \p U_2} \Psi_{\xi^n}(U_1, U_1, U_2) + 
C_+ {\p \over \p U_2} \Psi_{\xi^{M_+}}(U_1, U_1, U_2)} &=& 0 \ , \label{Ht17} \\
- \Psi_{\xi^n}(U_1, U_1, U_2) 
+ C_+ \Psi_{\xi^{M_+}}(U_1, U_1, U_2) && \no \\ 
- \Psi_{\xi^n}(U_2, U_1, U_2)  
+ C_+ \Psi_{\xi^{M_+}}(U_2, U_1, U_2) &=& 0 \ .
\label{Ht18}
\end{eqnarray}
These two equations have a solution $U_1=U_2=U^*$, where $U^*$ is
determined by
$$-\Psi_{\xi^n}(U^*, U^*, U^*) + C_+ \Psi_{\xi^{M_+}}(U^*, U^*, U^*) 
= {1 \over 2} {d \over d \xi} [ - \xi^n + C_+ \xi^{M_+} ]|_{\xi=U^*} = 0 \ ,$$
where we have used the boundary condition (\ref{BP2}) for $\Psi_{\xi^n}$ and $\Psi_{\xi^{M_+}}$.
We hence obtain
$$U_1=U_2=U^{*}=[{n \over C_+ M_+}]^{1 \over M_+ - n}$$
as a solution of (\ref{Ht17}-\ref{Ht18}).

Second, we calculate the Jacobian of equations (\ref{Ht15}) and (\ref{Ht16})
at $U_1=U_2=U^{*}$ and $T=0$. Denote the left hand side of 
(\ref{Ht15}) by $F_1(U_1,U_2,T)$ and that of (\ref{Ht16}) by
$F_2(U_1,U_2,T)$. Since $\Psi_g$ satisfies the Euler-Poisson-Darboux equations
(\ref{ph1}), ({\ref{ph2}) and (\ref{BP2}), we obtain
\begin{eqnarray*}
{\p F_1 \over \p U_1} = - {\p F_1 \over \p U_2} &=&
\sqrt{ - {\p \over \p U_2} \Psi_{\xi^n}(U^*, U^*, U^*) +
C_+ {\p \over \p U_2} \Psi_{\xi^{M_+}}(U^*, U^*, U^*)} \\
&=& \sqrt{{1 \over 8}
{d^2 \over d \xi^2} [- \xi^n + C_+ \xi^{M_+}]}|_{\xi=U^*}> 0 \ ,  \\
{\p F_2 \over \p U_1} = {\p F_2 \over \p U_2} &=& 4 [- {\p \over \p U_2}
\Psi_{\xi^n}(U^*, U^*, U^*) + C_+ {\p \over \p U_2} \Psi_{\xi^{M_+}}(U^*, U^*, U^*)] \\
&=& {1 \over 2} {d^2 \over d \xi^2} [- \xi^n + C_+ \xi^{M_+}]|_{\xi=U^*} > 0 \ . 
\end{eqnarray*} 
Hence, the Jacobian of (\ref{Ht15}) and (\ref{Ht16}) is nonzero. 
Equations (\ref{Ht15}) and (\ref{Ht16}) then give $U_1$ and $U_2$ as functions
of $T$ near $T=0$.

Therefore, equations (\ref{Ht11}-\ref{Ht12}) have a solution $u_1(t)$,
$u_2(t)$ for negatively large $t$. The solution has the following
asymptotics
\begin{equation}
u_1(t) \approx U^* |t|^{1 \over M_+ - n} \ , ~~~
u_2(t) \approx U^* |t|^{1 \over M_+ - n} ~~~~~~~~ \mbox{for $t \ll - 1$} \ .
\no
\end{equation}
Moreover, it follows from (\ref{Ht15}) and (\ref{U12}) that the length of 
the interval $[u_1(t), u_1(t)]$, the support of the minimizer, shrinks to
zero as $t \rightarrow - \infty$.

To make sure that $\psi(\xi)$ of (\ref{minimizer}) is the minimizer, we 
also need to verify that inequalities (\ref{constraint}-\ref{constraint2})
are satisfied.

We split $\Phi_0$ in the fashion of (\ref{split})
$$\Phi_0 = t \Phi_{\xi^n} + 
C_+ \Phi_{\xi^{M_+}} + \Phi_{h_+(\xi)} \ .$$
Using the scaling (\ref{U12}), we write $\Phi_0(\xi, u_1(t), u_2(t))$
as
\begin{equation}
\label{est}
|t|^{M_+ -1 \over M_+ -n} [- \Phi_{\xi^n}(\Xi,U_1,U_2)
+C_+ \Phi_{\xi^{M_+}}(\Xi,U_1,U_2) + {1 \over |t|^{M_+ -1 \over M_+ -n}} 
\Phi_{h_+(\xi)}(\xi, u_1, u_2)] \ ,
\end{equation}
where $\Xi = \xi /|t|^{1 \over M_+ -n}$. 
We use (\ref{V*hh}) to deduct that if $\Xi$ is kept bounded, say, 
$U^*/3 \leq \Xi \leq 3 U^*$, the last term in the  
parenthesis goes to zero uniformly as $t \rightarrow - \infty$. 
The sum in the parenthesis then has the limit
\begin{eqnarray}
\lefteqn{- \Phi_{\xi^n}(\Xi,U^*,U^*)+ C_+ \Phi_{\xi^{M_+}}(\Xi,U^*,U*)} \no \\  
&=& { {d \over d \Xi}[- \Xi^n + C_+ \Xi^{M_+}]
- {d \over d \xi}[- \xi^n + C_+ \xi^{M_+}]|_{\xi = U^*} \over 2(\Xi - U^*)} \no \\
&=& { {d \over d \Xi}[- \Xi^n + C_+ \Xi^{M_+}] \over 2(\Xi - U^*)} \no \\
&=& {C_+ {M_+} \Xi^{n-1} [ \Xi^{{M_+}-n} - U^{*^{{M_+}-n}}] \over 2(\Xi - U^*)} > 0  
\quad \mbox{for ${U^* \over 3} \leq \Xi \leq 3 U^*$} \ . \no  
\end{eqnarray}
Here we have used formula (\ref{H0u11}) for $\Phi_{\xi^n}$ and $\Phi_{\xi^{M_+}}$ in the 
first equality and 
$U^* = [n/C_+ M_+]^{1/(M_+ -n)}$ in the last two equalities.
In view of (\ref{est}), we have the inequality $\Phi_0(\xi, u_1(t), u_2(t)) > 0$ for  
${1 \over 2} u_2(t) \leq \xi \leq 2 u_1(t)$ when $t$ is negatively large. 
This proves the inequality
(\ref{constraint}).

To prove inequality (\ref{constraint2}), we first show that 
$\Phi_0(\xi, u_1(t), u_2(t)) > 0$ uniformly for $\xi < -1$ and $\xi > 2 u_1(t)$
when $t$ is negatively large. 

We shall prove the uniform positivity for $\xi < -1$
first. We use formula (\ref{H0u12}) to write
$\Phi_0(\xi, u_1(t), u_2(t))$ as
\begin{eqnarray}
\lefteqn{{1 \over 2 \sqrt{(\xi - u_1)(\xi - u_2)}} [V'(\xi) } \no \\
&&  - \ {1 \over \pi}
\int_{u_2}^{u_1} {\sqrt{(\xi - u_1)(\xi - u_2)} \over \xi - \mu} 
\frac{V'(\mu) d \mu}{\sqrt{(u_1 - \mu)(\mu - u_2)}} ] \no \\
&=& {1 \over 2 \sqrt{(\xi - u_1)(\xi - u_2)}} [V'(\xi) \no \\
&& - \ {1 \over \pi}
\int_{u_2}^{u_1} ({\sqrt{(\xi - u_1)(\xi - u_2)} \over \xi - \mu} - 1) 
\frac{V'(\mu) d \mu}{\sqrt{(u_1 - \mu)(\mu - u_2)}} ] \ , \label{est2} 
\end{eqnarray}
where in the last equality we have used equation (\ref{Ht14'}).

We now show that the integral of (\ref{est2}) tends to zero uniformly for $\xi < 0$ as
$t \rightarrow - \infty$. Because of the inequality 
$$| {\sqrt{(\xi - u_1)(\xi - u_2)} \over \xi - \mu} - 1 | \leq {u_1 - u_2 \over u_2 - \xi}
\leq {u_1 - u_2 \over u_2 }  \quad \mbox{for $\xi < 0 < u_2 < u_1$} \ ,$$
the integral is bounded by
$$ { u_1 - u_2 \over  u_2} \int_{u_2}^{u_1} {|V'(\mu)| \over \sqrt{(u_1 - \mu)(\mu - u_2)}}
d \mu $$
for $\xi < 0$.
This, in view of the scaling (\ref{U12}), is further bounded by a constant times
$$ { [U_1(t) - U_2(t)]^2 \over u_2(t)} |t|^{M_+ \over M_+  - n} \ ,$$
where we have used $V'(\mu) = O( (u_1(t) - u_2(t) ) t^{M_+ -2 \over M_+  -n})$
for $u_2(t) \leq \mu \leq u_1(t)$. This asymptotics can be verified by observing that $V'(\mu)$
has a zero between $u_2(t)$ and $u_1(t)$ because of (\ref{Ht14'}) and by expanding
$V'(\mu)$ around the zero. 
Since $U_1(t) - U_2(t) = O(|t|^{-{M_+ \over 2(M_+ -n)}})$ because of (\ref{Ht15}),    
the integral of (\ref{est2}) tends to zero uniformly for $\xi < 0$ as $t \rightarrow - \infty$.

The first term $V'(\xi) = t n \xi^{n-1} + V'_*(\xi) $ in the parenthesis of 
(\ref{est2}) is bounded from above by a negative constant
for $\xi \leq -1$ when $t$ is negatively large. This immediately follows from condition (\ref{V*-}).

Therefore, 
$\Phi_0(\xi, u_1(t), u_2(t)) > 0$ uniformly for $\xi < -1$ when $t$ is negatively large.

In the same way, we can prove that $\Phi_0(\xi, u_1(t), u_2(t)) > 0$ uniformly for
$\xi > 2 u_1(t)$ when $t$ is negatively large. This together with 
$\Phi_0(\xi, u_1(t), u_2(t)) > 0$ for ${1 \over 2} u_2(t) \leq \xi \leq 2 u_1(t)$
proves the first half of (\ref{constraint2}).

It remains to prove the rest of (\ref{constraint2}), i.e.,
\begin{equation}
\int_{\xi}^{u_2(t)} \sqrt{(\mu - u_1(t))(\mu - u_2(t))}
\ \Phi_0(\mu, u_1(t), u_2(t)) d \mu < 0
\label{in}
\end{equation}
for $\xi < u_2(t)$. Since $\Phi_0(\mu, u_1(t), u_2(t)) > 0$ uniformly for $\xi < -1$
and $\xi \geq u_2(t)/2$
when $t$ is negatively large, it suffices to prove the inequality for
$-1 \leq \xi \leq u_2(t)/2$.

Using formula (\ref{H0u12}) for $\Phi_0(\xi, u_1(t), u_2(t))$, we write the left hand
side of the inequality of (\ref{in}) as half of
\begin{eqnarray}
\lefteqn{- V(\xi) + V(u_2(t))} \label{est3} \\
& & - \ {1 \over \pi} \int_{\xi}^{u_2(t)} \sqrt{(\xi - u_1)(\xi - u_2)}
\ [\int_{u_2}^{u_1} \frac{V'(\mu) d \mu }{(\xi - \mu)\sqrt{(u_1 - \mu)(\mu - u_2)}} ] d \xi \ .
\no
\end{eqnarray}
The first two terms $- V(\xi) + V(u_2(t)) \leq - V(u_2(t)/2) + V(u_2(t) < 0$ for 
$-1 \leq \xi \leq u_2(t)/2$.
The whole third term, in view of the scaling (\ref{U12}), has an order in $t$
lower than that of $- V(u_2(t)/2) + V(u_2(t))$. These prove
that (\ref{est3}) is negative for $-1 \leq \xi \leq u_2(t)/2$ when $t$ is negatively
large. We have therefore proved (\ref{in}).  

Inequalities (\ref{constraint}-\ref{constraint2}) have thus been verified.
By Theorem 2.3, the equilibrium measure
for the external field $V(\xi) = t \xi^n + V_*(\xi)$
has no gap in its
support when $t$ is negatively large enough.

The scaling nature of the above approach allows us to extend the result
from $p(\xi) = \xi^n$ to an arbitrary monic polynomial of degree $n$.

\begin{theo}
Under conditions (\ref{V*}-\ref{V*h}) on $V_*(\xi)$, 
the equilibrium measure
for the external field $V(\xi,t) = V_*(\xi) + t \ p(\xi)$, where $p(\xi)$ is
a monic polynomial of an odd degree $n$, has no gap in its
support when $t$ is negatively large enough.
\end{theo}

The case when $t$ is positively large can be treated in the same way.
Here, we replace conditions (\ref{V*}-\ref{V*hh}) on $V_*(\xi)$ by analogous ones
\begin{eqnarray}
\label{V*2}
V_*(\xi) &=& C_{-} |\xi|^{M_-} + h_{-}(\xi) ~~~~~~~~ \mbox{ for $\xi \ll -1$} \ , \\
V_*'(\xi) &\geq& 0 \label{V*2-} ~~~~~~~~~~~~~~~~~~~~~~~~~~~~~ \mbox{ for $\xi \gg 1$} \ , 
\end{eqnarray}
where $C_->0$ and $M_-> max\{2, n\}$. Function $h_+(\xi)$ has the following behavior
\begin{equation}
\label{V*2h}
\lim_{\xi \rightarrow - \infty} {h_-'''(\xi) \over |\xi|^{M -3}} = 0  \ .
\end{equation}

\begin{theo}
Under conditions (\ref{V*2}-\ref{V*2h}) on $V_*(\xi)$,
the equilibrium measure
for the external field $V(\xi,t) = V_*(\xi) + t \ p(\xi)$, where $p(\xi)$ is
a monic polynomial of an odd degree $n$, has no gap in its
support when $t$ is positively large enough.
\end{theo}

\medskip

\noindent
{\bf $\S$ 3.2 \quad The degree of polynomial $p(\xi)$ is even}

\medskip

We shall show that the equilibrium measure for the $V_*(\xi) + t \ p(\xi)$ has 
a single gap in its support when $t$ is negatively large.
When $t$ is positively large, there are many possibilities.
The equilibrium measure has no gap when $p(\xi)$ is convex and
may have multi-gaps when $p(\xi)$ is non-convex.

We will first study the case when $t$ is negatively large. 
The minimizer will be supported on $[u_4, u_3] \cup [u_2, u_1]$.
This corresponds to $g=1$ in (\ref{support}).

For simplicity, we assume
that $V_*(\xi)$ of (\ref{V}) is an even function. The evenness of 
$V(\xi) = t \xi^n + V_*(\xi)$
implies the evenness of the equilibrium measure $\psi(\xi)$. Consequently,
its support must be symmetric about the origin. This means
that $u_3 = - u_2$ and $u_4 = - u_1$. We therefore obtain from (\ref{P}) that 
\begin{eqnarray*}
P(\xi, \vu) &=& 2(\xi^2 - u_1^2) (\xi^2 - u_2^2) \Phi_{1}(\xi, \vu) +
(\xi + u_1)(\xi^2 - u_2^2) \Psi_{1}(u_1, \vu) \\
&+& (\xi^2 - u_1^2)(\xi + u_2) \Psi_{1}(u_2, \vu)
 + (\xi^2 - u_1^2)(\xi - u_2) \Psi_{1}(- u_2, \vu) \\
&& + (\xi - u_1)(\xi^2 - u_2^2) \Psi_{1}(- u_1, \vu)
- {1 \over \pi} \xi \ .
\end{eqnarray*}
Equations (\ref{Hodo2}) exactly become
\begin{eqnarray*}
2u_1 (u_1^2 - u_2^2) \Psi_{1}(u_1, \vu)
- {1 \over \pi} u_1 &=& 0 \ , \\
2u_2 (u_2^2 - u_1^2) \Psi_{1}(u_2, \vu)
- {1 \over \pi} u_2 &=& 0 \ ,
\end{eqnarray*}
which are equivalent to
\begin{eqnarray}
(u_1 - u_2) \sqrt{2(u_1+u_2)} \sqrt{\p \Psi_{1}(u_1, \vu) \over \p u_2}
- \sqrt{1 \over \pi} &=& 0 \ , \label{even3} \\
\Psi_{1}(u_1, \vu) + \Psi_{1}(u_2, \vu) &=& 0 \label{even4}  \ .
\end{eqnarray}
We have used the following identity when deriving (\ref{even3})
$$\Psi_1(u_1, \vu) - \Psi_1(u_2, \vu) = 2(u_1 - u_2) {\p \over \p u_2} 
\Psi_1(u_1, \vu) \ .$$
Its proof is the same as the one for a similar identity (\ref{identity}).

In view of the integral formula (\ref{psi}) for its left hand side,
equation (\ref{even4}) is equivalent to
\begin{equation}
\label{even5'}
\int_{u_2}^{u_1} {V'(\mu)  d \mu \over \sqrt{(u_1^2 - \mu^2) (\mu^2 - u_2^2)}} = 0 \ .
\end{equation}

We further assume $V_*$ satisfies the condition
\begin{equation}
\label{V**}
V_*(\xi) = C |\xi|^M + h(\xi) ~~~~~~~~ \mbox{ for $|\xi| \gg 1$} \ ,
\end{equation}
where $C>0$ and $M> max \{3, n \}$. Here $h(\xi)$ satisfies
\begin{equation}
\label{V**h}
\lim_{\xi \rightarrow \pm  \infty} {h''''(\xi) \over \xi^{M -4}} = 0  \ .
\end{equation}
This implies
\begin{equation}
\label{V**hh}
\lim_{\xi \rightarrow \pm  \infty} {h(\xi) \over \xi^M} = 0  \ , ~
\lim_{\xi \rightarrow \pm  \infty} {h'(\xi) \over \xi^{M-1}} = 0  \ , ~
\lim_{\xi \rightarrow \pm  \infty} {h''(\xi) \over \xi^{M -2}} = 0  \ , ~
\lim_{\xi \rightarrow \pm  \infty} {h'''(\xi) \over \xi^{M -3}} = 0  \ .
\end{equation}

We will solve equations (\ref{even3}-\ref{even4}) for large $u_1$ and $u_2$.

Again, we will first consider $p(\xi) = \xi^n$.

We split $\Psi_1$ of (\ref{even3}-\ref{even4}) into simpler terms
\begin{equation}
\Psi_1 = t \Psi_{\xi^n} + C \Psi_{\xi^M} + h(\xi) \ ,
\no
\end{equation}
where $\Psi_{\xi^n}$ and $\Psi_{\xi^M}$ are homogeneous functions of $(\xi, u_1, u_2)$
of orders $n-2$ and $M-2$, respectively.

Introducing
\begin{equation}
\label{s5}
U_1 = {u_1 \over |t|^{1 \over M-n}} \ , \quad U_2 = {u_2 \over |t|^{1 \over M-n}} \ ,
\end{equation}
we scale equations (\ref{even3}-\ref{even4}) as
\begin{eqnarray}
\lefteqn{(U_1 - U_2) \sqrt{2 (U_1 + U_2)} \{ - {\p \over \p U_2}
\Psi_{\xi^n}(U_1, U_1, U_2, - U_2, -U_1)} \no \\
&& + C{\p \over \p U_2} \Psi_{\xi^M}(U_1, U_1, U_2, -U_2, -U_1) \label{even5} \\
& & + {1 \over |t|^{M-3 \over M-n}}
{\p \over \p u_2} \Psi_{h(\xi)}(u_1, u_1, u_2, -u_2, -u_1) \}^{1 \over 2} 
- {1 \over \sqrt{\pi} |t|^{M \over 2(M-n)}} = 0 \ , \no \\
\lefteqn{- \Psi_{\xi^n}(U_1, U_1, U_2, -U_2, -U_1) + C \Psi_{\xi^M}(U_1, U_1, U_2,-U_2,-U_1)} \no \\
&& + {1 \over |t|^{M-2 \over M-n}}  \Psi_{h(\xi)}(u_1, u_1, u_2,-u_2,-u_1)  
- \Psi_{\xi^n}(U_2, U_1, U_2,-U_2,-U_1) \label{even6} \\
&& + C \Psi_{\xi^M}(U_2, U_1, U_2,-U_2,-U_1) +
{1 \over |t|^{M-2 \over M-n}}  \Psi_{h(\xi)}(u_2, u_1, u_2,-u_2,-u_1) = 0
\ . \no
\end{eqnarray}

At $t= - \infty$, equations (\ref{even5}-\ref{even6}) become
\begin{eqnarray}
\lefteqn{(U_1 - U_2) \sqrt{2 (U_1 + U_2)} \{ - {\p \over \p U_2}
\Psi_{\xi^n}(U_1, U_1, U_2, - U_2, -U_1)} \no \\
&& + C
{\p \over \p U_2} \Psi_{\xi^M}(U_1, U_1, U_2, -U_2, -U_1) \}^{1 \over 2} = 0 \ , \label{even7} \\
\lefteqn{- \Psi_{\xi^n}(U_1, U_1, U_2, -U_2, -U_1) + C \Psi_{\xi^M}(U_1, U_1, U_2,-U_2,-U_1)} \no \\
&& - \Psi_{\xi^n}(U_2, U_1, U_2,-U_2,-U_1) 
+ C \Psi_{\xi^M}(U_2, U_1, U_2,-U_2,-U_1) = 0 \ .
\label{even8}
\end{eqnarray}
They have a solution 
$U_1 = U_2 = \hat{U}$, where $\hat{U}$ is defined by 
$$ - \Psi_{\xi^n}(\hat{U}, \hat{U}, \hat{U}, - \hat{U}, - \hat{U}) + 
C \Psi_{\xi^M}(\hat{U}, \hat{U}, \hat{U}, - \hat{U}, - \hat{U})
= {1 \over 4 \hat{U}} {d \over d \xi} [ - \xi^n + C \xi^M ]|_{\xi=\hat{U}} = 0 \ ,$$
where we have used formula (\ref{psi'}) for $\Psi_{\xi^n}$ and $\Psi_{\xi^M}$ in the first
equality. We hence obtain
$$U_1=U_2=\hat{U}=[{n \over CM}]^{1 \over M - n}$$
as a solution of (\ref{even7}-\ref{even8}).

It is straight forward to use the Implicit Function Theory to determine the solution of 
equations (\ref{even5}-\ref{even6}) in the neighborhood of $U_1 = U_2 = \hat{U}$,
$t= - \infty$. This in turn gives solution $u_1(t)$, $u_2(t)$ of (\ref{even3}-\ref{even4})
for $t \ll -1$. The solution has the asymptotics 
$$u_1(t) \approx \hat{U} |t|^{1 \over M - n} \ , \quad u_1(t) \approx \hat{U} |t|^{1 \over M - n} 
\quad \mbox{for $t \ll -1$} \ .$$

We now verify inequalities (\ref{constraint}-\ref{constraint2}). We will first prove
(\ref{constraint}) and the last two inequalities of (\ref{constraint2}). 
It suffices
to show that $\Phi_1(\xi, u_1(t), u_2(t), -u_2(t), -u_1(t))$ is negative for 
$\xi \leq - u_2(t)/2$ and positive for $\xi \geq u_2(t)/2$
when $t \ll -1$. Since $\Phi_1(\xi, u_1(t), u_2(t), -u_2(t), -u_1(t))$ is odd in $\xi$
in view of formula (\ref{H1u12}), we only need to prove its positivity for
$\xi \geq u_2(t)/2$.

We split $\Phi_1$ into simpler terms and scale these terms
\begin{eqnarray}
\lefteqn{\Phi_1(\xi, u_1, u_2, -u_2, -u_1)} \no \\
&=&|t|^{M-3 \over M-n} [- \Phi_{\xi^n}(\Xi,U_1,U_2,-U_2,-U_1)
+C \Phi_{\xi^M}(\Xi,U_1,U_2,-U_2,-U_1) \no \\
&& + {1 \over t^{M-3 \over M - n}} \Phi_{h(\xi)}(\xi,u_1,u_2,-u_2,-u_1)] \ ,
\label{est4}
\end{eqnarray}
where $\Xi = \xi /|t|^{1 \over M-n}$.
In view of condition (\ref{V**hh}) and the scaling (\ref{s5}), the last term in the 
parenthesis goes to zero uniformly for $\hat{U} /3 \leq \xi /|t|^{1 \over M-n} \leq
3 \hat{U}$ as $t \rightarrow - \infty$. The sum in the parenthesis then has the limit
\begin{eqnarray*}
\lefteqn{- \Phi_{\xi^n}(\Xi,
\hat{U},\hat{U}, -\hat{U}, -\hat{U}) + C \Phi_{\xi^M}(\Xi, \hat{U}, \hat{U},
- \hat{U}, - \hat{U})} \\
&=& { \hat{U} {d \over d \Xi}[- \Xi^n + C \Xi^M]
- \Xi {d \over d \xi}[- \xi^n + C \xi^M]|_{\xi = \hat{U}} \over 2 \hat{U}(\Xi^2-\hat{U}^2)} \\
&=& {CM\Xi^{n-1} [\Xi^{M-n} - \hat{U}^{M-n}] \over \Xi^2 - \hat{U}^2} \ ,
\end{eqnarray*}
where we have used (\ref{H1u11}) in the first equality and $\hat{U}=[{n / CM}]^{1 / M - n}$
in the last one. The limit is  
positive for 
$\hat{U} /3 \leq \Xi \leq 3 \hat{U}$.
It then follows from equation (\ref{est4}) that $\Phi_1(\xi,u_1(t),u_2(t),-u_2(t),-u_1(t))$
is positive for 
$u_2(t) / 2 \leq \xi \leq 2 u_1(t)$ when $t \ll -1$. 

We now show that 
$\Phi_1(\xi,u_1(t),u_2(t),-u_2(t),-u_1(t)) > 0$ uniformly for $\xi \geq  2 u_1(t)$.
We use formula (\ref{H1u12}) to write $\Phi_1(\xi, u_1, u_2, -u_2, -u_1)$ as
\begin{eqnarray}
\lefteqn{{\xi \over \sqrt{(\xi^2 - u_1^2)(\xi^2 - u_2^2)}}[ {V'(\xi) \over 2 \xi}} \no \\
&& + 
{ 1 \over  \pi} \int_{u_2}^{u_1} {\sqrt{(\mu^2 - u_1^2)(\mu^2 - u_2^2)} 
\over (\mu^2 - \xi^2)}{V'(\mu) d \mu  
\over \sqrt{(u_1^2 - \mu^2)(\mu^2 - u_2^2)}}] \no \\
&=& {\xi \over \sqrt{(\xi^2 - u_1^2)(\xi^2 - u_2^2)}} [ {V'(\xi) \over 2 \xi} \no \\
&& + { 1 \over  \pi} \int_{u_2}^{u_1} ({\sqrt{(\mu^2 - u_1^2)(\mu^2 - u_2^2)}
\over \mu^2 - \xi^2} - 1) {V'(\mu) d \mu \over 
\sqrt{(u_1^2 - \mu^2)(\mu^2 - u_2^2)}}] \ , \label{2} 
\end{eqnarray}
where we have used (\ref{even5'}) in the equality. 

It is then straight forward to use the argument below (\ref{est2}) to show that  
the integral of (\ref{2}) tends to zero uniformly for $\xi \geq 2 u_1(t)$
as $t \rightarrow - \infty$. 

The first term $V'(\xi) = tn \xi^{n-1} + V_*'(\xi)$ in the parenthesis is
bounded from below by a positive constant uniformly
for $\xi \geq  2 u_1(t)$ when $t \ll -1$. This follows from the conditions 
(\ref{V**}-\ref{V**hh}) on $V_*(\xi)$.

Function $\Phi_1(\xi,u_1(t),u_2(t),-u_2(t),-u_1(t))$ is therefore positive
in view of (\ref{2}).
This together with the similar result for 
$u_2(t) / 2 \leq \xi \leq 2 u_1(t)$ proves that it is positive uniformly for
$\xi \geq u_2(t)/2$ when $t \ll-1$.

We have therefore proved (\ref{constraint}) and the last two inequalities
of (\ref{constraint2}).  

It remains to prove the rest of (\ref{constraint2}), i.e.,
\begin{equation}
\label{ine5}
\int_{- u_2(t)}^{\xi} \sqrt{(\mu^2 - u_1(t)^2)(\mu^2 - u_2(t)^2)}
\ \Phi_1(\mu, u_1(t), u_2(t),-u_2(t),-u_1(t)) d \mu > 0  
\end{equation}
for $|\xi| < u_2(t)$
when $t \ll -1$. 

First, the inequality (\ref{ine5}) is valid for $- u_2(t) < \xi \leq - u_2(t)/2$ since
$\Phi_1$ is negative for $- 2 u_1(t) \leq \xi \leq - u_2(t) / 2$.
It suffices to prove the inequality for $- u_2(t) / 2 \leq \xi \leq u_2(t)$.

We use formula (\ref{H1u12})
to write the integral of (\ref{ine5}) as
\begin{eqnarray*}
\lefteqn{{ V(\xi) - V(-u_2) \over 2}} \\
&& + {1 \over \pi} \int_{- u_2}^{\xi} \sqrt{
(\nu^2 - u_1^2)(\nu^2 - u_2^2)} \ [\int_{u_2}^{u_1} {V'(\mu) d \mu \over (\mu^2 - \nu^2)
\sqrt{(u_1^2 - \mu^2)(\mu^2 - u_2^2)}} ] \ \nu d \nu \ . 
\end{eqnarray*}
It is easy to use the scaling (\ref{s5}) to show that the above is positive for
$- u_2(t)/2 < \xi \leq 0$; hence, inequality (\ref{ine5}) is verified for $-u_2(t) 
< \xi \leq 0$. Since $\Phi_1(\xi, u_1, u_2, -u_2, -u_1)$ 
is odd in $\xi$ in view of (\ref{H1u12}), the inequality (\ref{ine5}) can be extended
from $- u_2(t) < \xi \leq 0$ to $- u_2(t) < \xi \leq u_2(t)$. This completes
the verification of inequalities (\ref{constraint}-\ref{constraint2}).

By Theorem 2.3, the equilibrium measure
for the external field $V(\xi) = t \xi^n + V_*(\xi)$
has a single gap in its
support when $t$ is negatively large.

It is also easy to extend the result from $p(\xi) = \xi^n$ to any
even monic polynomial of degree $n$.

\begin{theo}
Suppose $V_*(\xi)$ is an even function that satisfies condition (\ref{V**}-\ref{V**h}).
The equilibrium measure
for the external field $V(\xi,t) = V_*(\xi) + t \ p(\xi)$, where $p(\xi)$
is an even monic polynomial of degree $n$,
has a single gap in its
support when $t$ is negatively large.
\end{theo}

For the case that $t$ is positively large, we assume
\begin{equation}
\label{con2}
V_*''(\xi) \geq 0 ~~ \quad \mbox{for $|\xi| \gg 1$} \ .
\end{equation}

Again, we first consider the case $p(\xi) = \xi^n$

Since $V_*''(\xi)$ is uniformly bounded in $|\xi| \leq A$ for some large constant $A$, we may
choose $t$ positively large enough so that $V(\xi) = t \xi^n
+ V_*(\xi)$ is a convex function for $|\xi| \geq \epsilon_0$, where $\epsilon_0$ is
a tiny positive number. It is well known that an everywhere convex external field
has an equilibrium measure whose support is a connected finite interval \cite{saf}.
Hence, if $V(\xi) = t \xi^n + V_*(\xi)$ is also convex in $|\xi| < \epsilon_0$
for large $t>0$, the
corresponding equilibrium measure has no gap in its support.

It is therefore interesting to note that, even if $V(\xi) = t \xi^n + V_*(\xi)$
is never convex in a small neighborhood of $\xi = 0$,
the equilibrium measure can still be shown to have no gap in its support
for large $t>0$.  

The approach is similar to one we use in $\S$ 3.1. We will present the proof
briefly.

We still need to solve equations (\ref{Ht11}-{\ref{Ht12}). Using
$$U_1 = u_1 t^{1 \over n} \ , ~~~ U_2 = u_2 t^{1 \over n} \ ,$$
we scale the equations as
\begin{equation}
(U_1 - U_2) \sqrt{ {\p \over \p U_2}
\Psi_{\xi^n}(U_1, U_1, U_2) + {1 \over t^{1 \over n}}
{\p \over \p u_2} \Psi_{V_*}(u_1, u_1, u_2)}  - \sqrt{1 \over \pi} = 0 \ , \label{even1} 
\end{equation}
\begin{eqnarray}
\lefteqn{\Psi_{\xi^n}(U_1, U_1, U_2)} \no \\
&&  + {1 \over t^{1 \over n}}\Psi_{V_*}(u_1, u_1, u_2)
+ \Psi_{\xi^n}(U_2, U_1, U_2) 
+ {1 \over t^{1 \over n}} \Psi_{V_*}(u_2, u_1, u_2) = 0 \ .
\label{even2}
\end{eqnarray}

At $t = + \infty$, these equations become
\begin{eqnarray*}
(U_1 - U_2) \sqrt{ {\p \over \p U_2}
\Psi_{\xi^n}(U_1, U_1, U_2)} -  \sqrt{1 \over \pi} &=& 0 \ , \\
\Psi_{\xi^n}(U_1, U_1, U_2) +
\Psi_{\xi^n}(U_2, U_1, U_2) &=& 0 \ .
\end{eqnarray*}
They have a solution $U_1 = \tilde{U}, U_2= - \tilde{U}$, where $\tilde{U}$ is defined by
$$2 \tilde{U} \sqrt{{\p \over \p U_2} \Psi_{\xi^n}(\tilde{U}, \tilde{U}, -\tilde{U})} - \sqrt{1 \over \pi} = 0 \ .$$
It is not hard to get from a formula of type (\ref{Cauchy}) for $\Psi_{\xi^n}$ that
$${\p \over \p U_2} \Psi_{\xi^n} (\tilde{U}, \tilde{U}, - \tilde{U}) = {1 \over 4} {(n-1)!! \over (n-2)!!} U^{n-2} \ .$$
We hence obtain
$$U_1 = \tilde{U} \ , \quad U_2= - \tilde{U} \ , \quad  \tilde{U} = [{(n-2)!! \over \pi (n-1)!!}]^{1 \over n} \ .$$
as a solution of equations (\ref{even1}-\ref{even2}) at $t= + \infty$.

One can then use the Implicit Function Theory to show that (\ref{even1}-\ref{even2})
give $U_1$ and $U_2$ as functions of $t$ near $t= + \infty$ and
$$ lim_{t \rightarrow + \infty} U_1(t) = \tilde{U} \ , \quad
   lim_{t \rightarrow + \infty} U_2(t) = - \tilde{U} \ .$$
Therefore, equations (\ref{Ht11}-{\ref{Ht12}) can be inverted to give $u_1(t)$ and
$u_2(t)$, which have the asymptotics
$$u_1(t) \approx {\tilde{U} \over t^{1 \over n}} \ , \quad
u_2(t) \approx - {\tilde{U} \over t^{1 \over n}}  \quad \mbox{as $t \gg 1$} \ . $$

To verify the inequalities (\ref{constraint}-\ref{constraint2}), it suffices to
prove that $\Phi_0(\xi, u_1(t), u_2(t)) > 0$ uniformly for all $\xi$ when
$t$ is sufficiently large.

We first decompose $\Phi_0$
\begin{eqnarray*}
\Phi_0(\xi, u_1(t), u_2(t)) &=& t \Phi_{\xi^n}(\xi, u_1(t), u_2(t)) +
\Phi_{V_*(\xi)}(\xi, u_1(t), u_2(t)) \\
&=& t^{2 \over n} [ \Phi_{\xi^n}({\xi \over t^{1 \over n}}, U_1(t), U_2(t))    
+ {1 \over t^{2 \over n}} \Phi_{V_*(\xi)}(\xi, u_1(t), u_2(t))] \ .
\end{eqnarray*}

The first term in the parenthesis is bigger than a positive constant for all $\xi$ and large $t$.
To see this, function $\Phi_{\xi^n}$, in view of formula (\ref{H}), can be written as a multiple integral 
of the second derivative
of $\xi^n$. Hence, when $\tilde{U}/2 \leq  U_1 \leq 2 \tilde{U}$ and $-2 \tilde{U} \leq  U_2 \leq - \tilde{U}/2$, 
the first term is bounded
from below by a positive constant for all $\xi / t^{1 / n}$. 

To show $\Phi_0(\xi, u_1(t), u_2(t)) > 0$ for all $\xi$, it is enough to
show that the second term $\Phi_{V_*(\xi)}$ is bounded from below 
for all $\xi$ and large $t$. To accomplish this, we deduct from condition (\ref{con2}) that 
$V_*''(\xi)$ is bounded from below
for all $\xi$. Function $\Phi_{V_*(\xi)}$, which can be written as a multiple integral of
$V_*''$ in view of formula (\ref{H}), is therefore bounded from below for all $\xi$, $u_1$ and
$u_2$.

We have therefore verified inequalities (\ref{constraint}-\ref{constraint2}).
By Theorem 2.3, the equilibrium measure for the external field $V(\xi,t) = V_*(\xi) + t \xi^n$ 
is supported on a single interval when $t \gg 1$.

To generalize $p(\xi)$ from $\xi^n$ to any convex polynomial $p(\xi)$, we notice that
the lowest order term $\xi^m$ in $p(\xi)$ must be of an even order and that its coefficient
must also be positive. The previous analysis centered around $\xi^n$ can be applied to
$\xi^m$.

\begin{theo}
Under condition (\ref{con2}), the equilibrium measure for the external field
$V(\xi,t) = V_*(\xi) + t \ p(\xi)$, where $p(\xi)$ is a convex polynomial,
has no gap in its support when $t \gg 1$.
\end{theo}

We conclude this section by making an observation on the case of non-convex $p(\xi)$.
Such a polynomial $p(\xi)$ has multiple ``wells''. They will be amplified and
become the ``wells'' of the external field $V(\xi,t) = V_*(\xi) + t \ p(\xi)$ as 
$t$ is positively large if $V_*(\xi)$ satisfies condition (\ref{con2}). Its equilibrium measure
is therefore likely to be supported on multiple disjoint intervals.

\bigskip

\renewcommand{\theequation}{A.\arabic{equation}}\setcounter{equation}{0}
\noindent
{\large \bf Appendix A. ~ Algebro-Geometric Solution of the Riemann-Hilbert Problem }

In Section 2, we use function theoretical methods to solve the Riemann-Hilbert problem 
(\ref{rh1}) and (\ref{rh2}). Our solution is given by formula (\ref{g}); it is the cornerstone of 
Propositions 2.1 and 2.2. In this appendix, we will present yet another approach
to the Riemann-Hilbert problem. We will solve it 
using an algebro-geometric method. More precisely, we will give another expression for
the solution (\ref{g}) and hence re-derive the formulae of Propositions 2.1 and 2.2. 

The Riemann-Hilbert problem (\ref{rh1}) and (\ref{rh2}) has an intrinsic algebro-geometric 
structure. All its solutions are connected to the Riemann surface defined by the 
equation $w^2= (\mu - u_1)(\mu - u_2) \cdots (\mu - u_{2g+2})$. We choose the
branch cuts along the set of $I$ of (\ref{support}). It is remarkable that our 
algebro-geometric approach does not require the external field $V(\xi)$ to be an analytic function. 

Our starting point is to generalize the Cauchy kernel.
On a Riemann surface, there are many analogues of the 
Cauchy kernel. The most convenient one for the Riemann-Hilbert 
problem (\ref{rh1}) and (\ref{rh2})  is  an Abelian differential
of the third kind, denoted by $K(\mu,z)d\mu$,
with two simple poles at the
points $(z, \pm R(z, \vu))$ with residues $\pm 1$, respectively. 
Here, $\pm R(z, \vu)$ are the upper and lower sheets of the Riemann surface. 
This means that $K(\mu,z)d\mu$ takes the form
\[
K(\mu,z)d\mu= \dfrac{d\mu}{R(\mu,\vu)}\dfrac{R(z,\vu)}{\mu-z}+
\mbox{holomorphic terms}
\]
and that its behavior, as $\mu$ is near the poles, is 
$$K(\mu,z)d\mu = \pm \dfrac{d\mu}{\mu-z} + \mbox{regular terms}. $$
We may further require that 
\begin{equation}
\label{loop2}
\int_{u_{2k+1}}^{u_{2k}} K(\mu,z)d\mu = 0 ~~~~~~~~~ k=1, 2, \cdots, g \ , 
\end{equation}
which is analogous to condition (\ref{loop}).

The differential $K(\mu,z)d\mu$ then takes the form
\begin{equation*}
K(\mu,z)d\mu=\dfrac{d\mu}{R(\mu,\vu)}\dfrac{R(z,\vu)}{\mu-z}+\sum_{k=1}^g\omega_k(\mu)d\mu
\int_{u_{2k+1}}^{u_{2k}}\dfrac{d\xi}{R(\xi,\vu)}\dfrac{R(z,\vu)}{\xi-z} \ ,
\end{equation*}
where $\omega_k(\mu)d\mu$ is the basis of holomorphic differentials  normalized
along the intervals $[u_{2k+1},u_{2k}]$, $k=1,\dots,g$.

Using the Riemann bilinear relations between $K(\mu,z)d\mu$  and
the differentials $\dfrac{P_{g,n}(\eta,\vu)}{R(\eta,\vu)}d\eta$ defined in 
(\ref{Pgn}), it is possible to reduce the above formula to the form
\begin{equation}
\label{omega}
K(\mu,z)d\mu=\dfrac{d\mu}{R(\mu,\vu)}\dfrac{R(z,\vu)}{\mu-z}+
\dfrac{1}{2}
\sum_{m=1}^g\dfrac{\mu^{g-m}d\mu}{R(\mu,\vu)}\sum_{k=1}^m k\Gamma_{m-k}(\vu) 
\int\limits_{p_-}^{p_+}\dfrac{P_{g,k}(\eta,\vu)}{R(\eta,\vu)}d\eta \ ,
\end{equation}
where  $p_{\pm}=(z,\pm R(z,\vu))$  and $\Gamma_k(\vu)$ are the coefficients 
of the expansion  (\ref{gamma}). Formula (\ref{omega}) can also be derived 
from the explicit form of  the Bergmann kernel on the Riemann surface  $w^2=R^2(\xi,\vu)$ \cite{BEL}.

We next point out a remarkable symmetry property
\[
\dfrac{\partial}{\partial z} K(\mu,z)=\dfrac{\partial}{\partial \mu} 
K(z,\mu) \ .
\]
It also follows from the Riemann bilinear relations \cite{Zv}.

We now use $K(\mu,z)d\mu$ to construct a solution of the Riemann-Hilbert problem
(\ref{rh1}) and (\ref{rh2}) 
\begin{align}
\label{par1}
\mathcal{G}(z)&=-\dfrac{1}{\pi}\int_I V'(\mu) K(z,\mu)d\mu+\dfrac{\sqrt{-1}}{\pi}\dfrac{P_{g,0}(z,\vu)}{R(z,\vu)}\\
\label{par2}
&=\dfrac{\partial}{\partial z} \left[\dfrac{1}{\pi}\int_I V(\mu) K(\mu,z)d\mu \right] + 
\dfrac{\sqrt{-1}}{\pi}\dfrac{P_{g,0}(z,\vu)}{R(z,\vu)} \ .
\end{align}
To see this, we derive from (\ref{par1}) the boundary value of $\mathcal{G}$ at real $\xi$   
\[
\mathcal{G}^+(\xi)= \sqrt{-1} V'(\xi)-\dfrac{1}{\pi} P.V.\int_I V'(\mu) K(\xi,\mu)d\mu
+\dfrac{\sqrt{-1}}{\pi}\dfrac{P_{g,0}(\xi)}
{R(\xi,\vu)} \ .
\]
When $\xi \in I$ and $\mu\in I$, the kernel $K(\xi,\mu)$ is real and $R(\xi,\vu)$ is pure imaginary.
We then derive $Im\,\mathcal{G}^+(\xi)=V'(\xi)$, which is (\ref{rh1}).
When $\xi \in \mathbb{R}\backslash \bar{I}$ and $\mu\in I$, the kernel  
$K(\xi,\mu)$ is pure imaginary and $R(\xi,\vu)$ is real.
We instead have  $Re\,\mathcal{G}^+(\xi)=0$, which is exactly (\ref{rh2}).

We further claim that $\mathcal{G}$ of (\ref{par1}) equals the function defined in (\ref{g}).
To see this, it suffices to prove that $\mathcal{G}$ of (\ref{par1}) satisfies both the loop conditions
(\ref{loop}) and the asymptotics $\mathcal{G}(z)= - {1 \over \sqrt{-1} \pi z} + O(1/z^2)$ for large $|z|$.
Conditions (\ref{loop}) are easily verified, in view of similar conditions (\ref{loop2}) on $K(\mu,z)d\mu$
and (\ref{lc}) on $P_{g,0}$.  For large $|z|$, the kernel $K(z,\mu)$ behaves like
$O(\frac{1}{z^2})$. Indeed the first term of  $K(z,\mu)$
is proportional to $\frac{1}{R(z,\vu)(z-\mu)}$, which clearly decays at least as fast as 
 $\frac{1}{z^2}$ as $|z| \rightarrow  \infty$. The other terms of  $K(z,\mu)$ are of the form
 $\frac{z^{g-m}d\mu}{R(z,\vu)}=O(\frac{1}{z^{m+1}})$ for $|z|$ large and 
 $m=1,\dots,g$. 
This, together with the behavior (\ref{Pgn}) of
$P_{g,0}$, justifies the asymptotics $\mathcal{G}(z)= - {1 \over \sqrt{-1} \pi z} + O(1/z^2)$ for large $|z|$.

Inserting the explicit form (\ref{omega}) into the  
expression (\ref{par2}), we obtain 
\begin{eqnarray}
\mathcal{G}(z) &=&  {\partial \over \partial z} \left[ {R(z,\vu) \over \pi} \int_I 
{V(\mu) d \mu \over R(\mu,\vu)(\mu-z)} \right]  \label{par3} \\
&& + \ 2 \sqrt{-1} \sum_{m=1}^g 
q_{g,m}(\vu) \sum_{k=1}^m k
\Gamma_{m-k}(\vu) {P_{g,k}(z,\vu) \over R(z,\vu)}   
+ {\sqrt{-1} \over \pi}{P_{g,0}(z,\vu) \over R(z,\vu)} \ , \no
\end{eqnarray}
where $q_{g,m}(\vu)$ is given in (\ref{qk}).

The boundary value of $\mathcal{G}(z)$ of (\ref{par3}) on the real axis can be obtained by
observing that the first term has the boundary value at real $\xi$
$${\p \over \p \xi} \left[ -2 \sqrt{-1} R(\xi, \vu) \Psi_g(\xi, \vu) + \sqrt{-1} V(\xi) \right] \ ,$$
where $\Psi_g$ is as given in Proposition 2.2. 
The function $\Psi_g(\xi,\vu)$ is related to 
$\Phi_g(z,\vu)$ of Proposition 2.1 by an 
identity \cite{GT} 
\begin{equation}
\label{r1}
\Phi_g(\xi,\vu)={\p \over \p \xi} \Psi_g(\xi,\vu)+\dfrac{1}{2}\sum_{i=1}^{2g+2}
\dfrac{\Psi_g(\xi,\vu)-\Psi_g(u_i,\vu)}{\xi-u_i}.
\end{equation}
We hence arrive at the boundary value of $\mathcal{G}$ at the real $\xi$
\[
\mathcal{G}^+(\xi)=\dfrac{ -2 \sqrt{-1} R^2(\xi,\vu) \Phi_g(\xi,\vu)
+ \sqrt{-1} Q(\xi)}{R(\xi,\vu)} + \sqrt{-1} V'(\xi) \ ,
\]
where the polynomial $Q(\xi,\vu)$ is
\[
-R^2(\xi,\vu)\sum_{i=1}^{2g+2}\dfrac{\Psi_g(u_i,\vu)}{\xi-u_i} +
2\sum_{m=1}^g q_{k,m}(\vu)\sum_{k=1}^m k
\Gamma_{m-k}(\vu) P_{g,k}(\xi,\vu)+\dfrac{1}{\pi}P_{g,0}(\xi,\vu).
\]
These are equivalent to the formulae of Propositions 2.1 and 2.2.

\bigskip

\renewcommand{\theequation}{B.\arabic{equation}}\setcounter{equation}{0}
\noindent
{\large \bf Appendix B. ~ Euler-Poisson-Darboux Equations}

The boundary value problem for the Euler-Poisson-Darboux equations
(\ref{ph1}-\ref{BP})
has one and only one solution. Its solution
can be constructed using those of the following simpler problem as building
blocks \cite{Tian1, GT}
\begin{eqnarray}
\label{EPD}
2(x_{1} - x_{2})\frac{\partial^{2} q}{\partial x_{1} \partial
x_{2}} & = & \frac{\partial q}{\partial x_{1}} - \rho \frac{\partial q}
{\partial x_{2}} \ , ~~~ \rho > 0  ~ is ~ a ~ constant \ , \\
q(x_1, x_1) & = & g(x_1) \ . \label{EPDd}
\end{eqnarray}

A simple calculation shows that the solution of (\ref{EPD}-\ref{EPDd}) is
given by the formula \cite{Tian1}
\begin{displaymath}
q(x_1, x_2) = C_0 \int_{-1}^{1} \frac{g(\frac{1+\mu}{2}x_1 + \frac{1-\mu}{2}x_2)}{
\sqrt{1 - \mu^{2}}} \ (1 + \mu)^{\frac{\rho - 1}{2}} d \mu \ ,
\end{displaymath}
where
\begin{displaymath}
C_0 = \frac{1}{\int_{-1}^{1} \frac{(1+\mu)^{\frac{\rho - 1}{2}}}{\sqrt{1-\mu^{2}}}
d \mu} \ .
\end{displaymath}
A change of integration variable gives another
formula for the solution
\begin{equation}
\label{EPQsS}
q(x_1,x_2) =   C_0 \left[{2 \over x_1-x_2}\right]^{\rho - 1 \over 2} \int_{x_2}
^{x_1} g(x) {(x - x_2)^{\rho - 2 \over 2} \over (x_1 - x)^{1 \over 2}} d x \ ,
\end{equation}
where the square root is set to be positive for $x$ between $x_1$ and $x_2$.

Using a solution method of \cite{Tian1},
one is able to construct the solution of (\ref{ph1}-\ref{BP}) using a multiple
integral\cite{GT}
\begin{equation}
\Phi_g(\xi, \vec{u}) =
M_0 \int_{-1}^{1} \cdots \int_{-1}^{1} \tilde{f}
\ {\prod_{k=2}^{2g+2} (1+\mu_k)^{k-1 \over 2}
\over \sqrt{ \prod_{k=1}^{2g+2} (1-\mu_k)}}
\ d \mu_1 \cdots d \mu_{2g+2} \ , \label{H}
\end{equation}
where $\tilde{f}$ denotes
\begin{displaymath}
f^{(g+2)}({1 + \mu_{2g+2} \over 2} (\cdots ({1 + \mu_{2} \over 2}
({1 + \mu_{1} \over 2} \xi + {1 - \mu_{1} \over 2}u_1) + \cdots ) +
{1 - \mu_{2g+2} \over 2} u_{2g+2} )
\end{displaymath}
and the constant $M_0$ is chosen so that the boundary condition (\ref{BP}) is
satisfied. Here $f^{(g+2)}$ denotes the $(g+2)th$ derivative of $f$. 

The function $\Psi_g(\xi, \vec{u})$ satisfies the same equations (\ref{ph1}-\ref{ph2})
as $\Phi_g(\xi, \vec{u})$; but, instead of the boundary condition (\ref{BP}),
it satisfies (\ref{BP2}) with one derivative lower.

We now list some of the properties concerning $\Phi_0$, $\Phi_1$ and $\Psi_1$. They are useful 
in the calculations in Section 3.

\begin{pb}

\begin{eqnarray}
\Phi_0(\xi, u, u) &=& { V'(\xi) - V'(u) \over 2(\xi - u)} \ , \label{H0u11} \\
\Phi_0(\xi, u_1, u_2) &=& {V'(\xi) \over 2\sqrt{(\xi - u_1)(\xi - u_2)}} \no \\
& &  - {1 \over 2\pi} \int_{u_2}^{u_1} 
{V'(\mu) d \mu \over (\xi - u) \sqrt{(u_1 - u)(u - u_2)}}  \ , \label{H0u12} \\
\Phi_1(\xi, u,u,-u,-u) &=& {u V'(\xi) - \xi V'(u) \over 2(\xi^2 - u^2)u} \ , \label{H1u11} \\
\Phi_1(\xi, u_1, u_2, -u_2, -u_1) &=& {V'(\xi) \over 2\sqrt{(\xi^2 - u_1^2)(\xi^2 - u_2^2)}} 
\no \\
& & + {\xi \over \pi} \int_{u_2}^{u_1} {V'(\mu) d \mu \over (\mu^2 - \xi^2) 
\sqrt{(u_1^2 - \mu^2)(\mu^2 - u_2^2)}} \ ,
\label{H1u12} \\
\Psi_1(u,u,u,-u,-u) &=& {V'(u) \over 4 u} \ , \label{psi'} \\ 
\Psi_1(u_1, u_1, u_2, -u_2, -u_1) &+& \Psi_1(u_2, u_1, u_2, -u_2, -u_1) \no \\
&=& {1 \over 2} \int_{u_2}^{u_1} {V'(\mu) d \mu \over 
\sqrt{(u_1^2 - \mu^2)(\mu^2 - u_2^2)}} \ . \label{psi}
\end{eqnarray}
Function $V(\xi)$ is assumed to be an even function in (\ref{H1u11}), (\ref{H1u12}), (\ref{psi'}) and (\ref{psi}).
\end{pb}

\proof  In formula (\ref{H}), $\Phi_0$, $\Phi_1$ and $\Psi_1$ are written as multiple 
integrals of $V''(\xi)$, 
$V'''(\xi)$ and $V''(\xi)$, respectively. Since the smooth function $V(\xi)$ can be approximated by
polynomials in $C^3(S)$ on every compact set $S$ as close as possible, it suffices to prove
Property B.1 when $V(\xi)$ is simply a polynomial.

We will rely on the contour integral formulation of $\Phi_g$
\begin{equation}
\label{Cauchy}
\Phi_g(\xi, \vu) = {1 \over 4 \pi \sqrt{-1}} \oint_{\gamma} {V'(\mu) d \mu \over (\mu - \xi)R(\mu, \vu)} \ ,
\end{equation}
where $\gamma$ is a contour enclosing the point $\xi$ and all the cuts along
$[u_{2k}, u_{2k-1}]$, $k=1, 2, \cdots, g+1$. To see this , it is easy to check that the right hand
side, as a function of $\xi$ and $\vu$, satisfies (\ref{ph1}) and (\ref{ph2}).
Boundary condition (\ref{BP}) is also easily verified using the Cauchy Integral Formula.

Identity (\ref{H0u11}) is an easy consequence of (\ref{Cauchy}) when $u_1=u_2=u$.
 
To prove (\ref{H0u12}), we replace the contour $\gamma$ by $\gamma'$, which still encloses the cuts
$[u_2, u_1]$, $k=1, 2, \cdots, g+1$, but excludes the point $\xi$, and write
$\Phi_0(\xi, u_1, u_2)$ as
$$ {1 \over 2} Res_{\mu = \xi} \left[{V'(\mu) \over (\mu - \xi) \sqrt{(\mu-u_1)(\mu - u_2)}} \right]
+ {1 \over 4 \pi \sqrt{-1}} \oint_{\gamma'} {V'(\mu) d \mu \over (\mu - \xi)R(\mu, \vu)} \ .$$
The first term gives the first term on the right of (\ref{H0u12}).
Rewriting the second term as an integral along the cut $[u_2, u_1]$ gives the second term of 
(\ref{H0u12}).

Identities (\ref{H1u11}) and (\ref{H1u12}) can be proved in the same way. 

To prove (\ref{psi'}), we use an analogous formula of (\ref{Cauchy}) for $\Psi_1(\xi,u_1,u_2,-u_2,-u_1)$
\begin{equation}
\label{Cauchy2}
{1 \over 4 \pi \sqrt{-1}} \oint_{\gamma} {V(\mu) d \mu \over (\mu - \xi)
\sqrt{(\mu^2 - u_1^2)(\mu^2 - u_2^2)}} \ .
\end{equation}
Identity (\ref{psi'}) is an immediate consequence of this formula and the Cauchy Integral Formula. 

To prove (\ref{psi}), we notice from the formula of type (\ref{H}) that $\Psi_1$ can be written as
a multiple integral of $V''(\xi)$. Since $V(\xi)$ is an even function of $\xi$, so is
$\Psi_1(\xi,u_1,u_2,-u_2,-u_1)$. This evenness allows us to modify formula (\ref{Cauchy2}) to get
a new contour integral
formulation for $\Psi(\xi,u_1,u_2,-u_2,-u_1)$
$${1 \over 4 \pi \sqrt{-1}} \oint_{\gamma} {V(\mu) \mu d \mu \over (\mu^2 - \xi^2)\sqrt{(\mu^2 - u_1^2)
(\mu^2 - u_2^2)}} \ .$$
The left hand side of (\ref{psi}) then equals 
\begin{eqnarray*}
\lefteqn{{1 \over 4 \pi \sqrt{-1}} \oint_{\gamma} \left( {1 \over \mu^2 - u_1^2} + {1 \over \mu^2 - u_2^2} \right)
{V(\mu) \mu d \mu \over 
\sqrt{(\mu^2 - u_1^2)(\mu^2 - u_2^2)}}} \\
&=& -{1 \over 4 \pi \sqrt{-1}} \oint_{\gamma} {\p \over \p \mu} \left( {1 \over \sqrt{(\mu^2 - u_1^2)(\mu^2 - u_2^2)}} \right) 
V(\mu) d \mu \\
&=& {1 \over 4 \pi \sqrt{-1}} \oint_{\gamma} {V'(\mu) d \mu \over \sqrt{(\mu^2 - u_1^2)(\mu^2 - u_2^2)}} \ .
\end{eqnarray*}
We obtain the right hand side of (\ref{psi}) by writing the last integral along the cuts
$[-u_1, -u_2]$ and $[u_2, u_1]$ and using the oddness of $V'(\mu)$. 

The proof of Property B.1 is completed.

{\bf Acknowledgments.} T. G. was supported in part by MISGAM ESF  Programme and ENIGMA MRTN-CT-2004-5652.
F.-R. T. did part of his research while he was visiting the Courant Institute in 2003,
IMS of CUHK and SISSA in 2004. F.-R. T. is grateful to these institutions for their support.
F.-R. T. was also supported in part by NSF Grant DMS-0103849 and Grant DMS-0404931 and by a 
John Simon Guggenheim Fellowship.


\begin{thebibliography}{50}

\bibitem{Zu} Bessis, D.; Itzykson, C.; Zuber, J.B. Quantum field theory techniques in graphical enumeration.
Adv. in Appl. Math.  1  (1980), 109--157.

\bibitem{bre} Brezin, E.; Itzykson, C.; Parisi, G.; Zuber, J.B.
Planar diagrams.
{\em Comm. Math. Phys.} 59 (1978), 35-51.

\bibitem{BR} 
Buyarov, V.S.; Rakhmanov, E.A.
On families of measures that are balanced in the external field on the real axis. 
{\em Sb. Math.} 190 (1999), 791-802.

\bibitem{BEL} Buchstaber, V.M.; Enolski\u\i, V. Z.; Le\u\i kin, D. V. Hyperelliptic Kleinian 
functions and applications.  {\em Solitons, geometry, and topology: on the crossroad},  1--33, 
Amer. Math. Soc. Transl. Ser. 2, 179, Amer. Math. Soc., Providence, RI, 1997. 

\bibitem{che} Chen, Y.; Grava, T. Eigenvalue correlations on 
hyperelliptic Riemann surfaces. {\em J. Phys. A} 35 (2002), L45-L49.

\bibitem{dei0} Deift, P. {\em Orthogonal polynomials and random matrices: A Riemann-Hilbert approach}.
AMS, Providence, 2000. 

\bibitem{dei1} Deift, P.; Kriecherbauer, T.; McLaughlin, K. T.-R. New
results on the equilibrium measure for logarithmic potentials in the presence
of an external field. {\em J. Approx. Theory} 95 (1998), 388-475.

\bibitem{dei2} Deift, P.; Kriecherbauer, T.; McLaughlin, K. T.-R.; Venakides, S.;
Zhou, X.
Strong asymptotics of orthogonal polynomials with respect to
exponential weights.  {\em Comm. Pure Appl.} Math. 52 (1999),
1491-1552.

\bibitem{dij}
Dijkgraaf, R.; Vafa, C. Matrix models, topological strings, and 
supersymmetric gauge theories.
{\em Nucl.Phys.} B644 (2002), 3-20.

\bibitem{dub} Dubrovin, B.A.
Geometry of 2D topological field theories. {\em Integrable Systems
and Quantum Groups}, Lecture Notes in Math. 1620 (1996), 120-348.


\bibitem{Ga} Gakhov, F.D. {\it Boundary value problems}, 
translated from the Russian. Reprint of the 1966 translation. 
Dover Publications, Inc., New York, 1990.

\bibitem{GT} Grava, T.; Tian, F.-R. The
generation, propagation and extinction of multiphases in the KdV zero 
dispersion limit.  {\em Comm. Pure Appl. Math.} 55 (2002), 1569-1639.

\bibitem{jur} Jurkiewicz, J. Regularisation of one-matrix models.
{\em Phys. Lett. B} 245 (1990), 178-84.

\bibitem{KM1} Kuijlaars, A.B.J.; McLaughlin, K.T.-R. Generic behavior of 
the density of states in random matrix theory and equilibrium problems 
in the presence of real analytic external fields.  {\em Comm. Pure Appl. 
Math.}  53 (2000), 736-785. 
 
\bibitem{lax} Lax, P.D.; Levermore, C.D. The small dispersion
limit for the Korteweg-de Vries equation I, II, and III. {\em Comm. Pure Appl.
Math. }
36 (1983), 253-290, 571-593, 809-830.

\bibitem{lax2} Lax, P.D.; Levermore, C.D.; Venakides, S. The generation and
propagation of oscillations in dispersive IVPs and their limiting behavior.
{\em Important developments in soliton theory 1980-1990}, 205-241.
Springer Series in Nonlinear Dynamics. 
Springer, Berlin (1993).

\bibitem{LS} Lubinsky, D.S.; Saff, E.B. {\em Strong asymptotics for extremal 
polynomials associated with weights on $\rm R$}. Lecture Notes in Math. 1305, Springer, Berlin, 1988.  

\bibitem{meh} Mehta, M.L. {\em Random matrices}. second
edition, Academic
Press, Boston, 1991.

\bibitem{Rn} 
Ransford, T. {\em Potential theory in the complex plane}. Cambridge University Press, Cambridge, UK, 1995.

\bibitem{saf} Saff, E.B.; Totik, V. {\em Logarithmic potential
with
external fields}. Springer, Berlin, 1997.

\bibitem{Tian1} 
Tian, F.-R. Oscillations of the zero dispersion
limit of
the
Korteweg-de Vries
equation. {\em Comm. Pure Appl. Math}. 46
(1993),
1093-1129.

\bibitem{Tian2} Tian, F.-R. The initial value problem for the Whitham averaged
system. {\em Comm. Math. Phys.} 166 (1994), 79-115.

\bibitem{ven} Venakides, S.
Higher order Lax-Levermore
theory.
{\em Comm.
Pure Appl. Math.} 43 (1990),
335-362.

\bibitem{Zv} Zverovich, E.I. Boundary value problems in the
theory of analytic functions in the H\"older classes on Riemann surfaces,
{\it Russ. Math. Survey,} 26 (1971), 118-185. 

\end{thebibliography}
\end{document}